\documentclass[a4paper]{jpconf}
\usepackage{graphicx}
\begin{document}

\newcommand{\NewA}{{\it New Astron.\/} }
\newcommand{\Nature}{{\it Nature\/} }
\newcommand{\ApJ}{{\it Astrophys. J.\/} }
\newcommand{\ApJS}{{\it Astrophys. J. Suppl.\/} }
\newcommand{\MNRAS}{{\it Mon. Not. R. Astron. Soc.\/} }
\newcommand{\PhRv}{{\it Phys. Rev.\/} }
\newcommand{\PhL}{{\it Phys. Lett.\/} }
\newcommand{\JCAP}{{\it J. Cosmol. Astropart. Phys.\/} }
\newcommand{\AeA}{{\it Astronom. Astrophys.\/} }
\newcommand{\etall}{{\it et al.\/} }
\newcommand{\arXiv}{{\it arXiv: \/} }

\def \apj  {ApJ}
\def \apjl  {ApJL}
\def \apjs  {ApJS}
\def \prd {Phy.Rev.D}
\def \mnras {MNRAS}
\def \hMsun {\ifmmode h^{-1}\,\rm M_{\odot} \else $h^{-1}\,\rm M_{\odot}$ \fi}
\def \hMpc {\ifmmode h^{-1}\,\rm Mpc \else $h^{-1}\,\rm Mpc$ \fi}
\def \Mpch {\ifmmode h\ {\rm Mpc}^{-1} \else $h\ {\rm Mpc}^{-1}$ \fi}
\def \hkpc {\ifmmode h^{-1}\,\rm kpc \else $h^{-1}\,\rm kpc$ \fi}
\def \LCDM {\ifmmode \Lambda{\rm CDM} \else $\Lambda{\rm CDM}$ \fi}

\title{High precision spectra at large redshift for dynamical DE cosmologies}

\author{S.A.~Bonometto, L.~Casarini}

\address{Department of Physics G.~Occhialini -- Milano--Bicocca
 University, Piazza della Scienza 3, 20126 Milano, Italy 
\& I.N.F.N., Sezione di Milano}

\author{A.V.~Macci\`o}

\address{Max-Planck-Institut f\"ur Astronomie, K\"onigstuhl 17, 69117
 Heidelberg}

\author{G.~Stinson}

\address{Jeremiah Horrocks Institute, University of Central
Lancashire, Presto PR1 2HE}

\begin{abstract}
The next generation mass probes will investigate DE nature by
measuring non--linear power spectra at various $z$, and comparing them
with high precision simulations. Producing a complete set of them,
taking into account baryon physics and for any DE state equation
$w(z)$, would really be numerically expensive. Regularities reducing
such duty are essential. This paper presents further n--body tests of
a relation we found, linking models with DE state parameter $w(z)$ to
const.--$w$ models, and also tests the relation in hydro simulations.
\end{abstract}

\section{Introduction}
The background metrics of space--time can be written in the forms

\vskip  .1truecm
$~~~~~~~~~~~~~~~~~~~~~~~~~~~~~~~
ds^2 = c^2 dt^2 - a^2(t)\, d\ell^2 = a^2(\tau)\, (d\tau^2 - d\ell^2)
$

\vskip .1truecm
\noindent
defining ordinary or conformal time, $t$ or $\tau$. In systems which
abandoned cosmic expansion, the former expression yields the
Minkowskian metrics: the no longer growing scale factor can be
inglobated in spatial coordinates. In such systems $t$ is the natural
time coordinate; e.g., many cyclic motions become (quasi--)periodical
in respect to $t$. On the contrary, when a system resent the overall
expansion, the natural time coordinate is $\tau$. For instance, sonic
waves in the pre--recombination plasma, are (quasi--)periodical in
respect to $\tau$. This periodicity yields the setting of maxima and
minima in CMB spectra, as well as the scales of BAO's in matter
spectra.

Let us consider then the comoving distance between today's observer
and the LSB (last scattering band), $L = \int_{rec}^0 d\ell$. It
coincides with the conformal time $T = \int_{rec}^0 d\tau$ a photon
takes to reach the observer from the LSB.  Accordingly, if two models
have the same comoving distance from the LSB, we expect them to share
those features which are determined by the cosmological dynamics. As a
matter of fact, it has been known since several years \cite{LJ} that,
if they share $\Omega_{c,b}$, $h$ and $n_s$ (density parameters,
Hubble parameter in units of 100$\, $(km/s)/Mpc, primeval scalar
index), their linear spectra are quite close. In particular, models
with the same conformal age, differing just because of DE state
equation, exhibit almost coincident linear spectra.

When non--linear effects are taken into account, the dynamics is no
longer simply ruled by $\tau$. Non linear spectral deformations arise
when matter halos form, abandoning the overall expansion. The point is
whether they are so widely spread to yield a significantly discrepant
spectral behavior at a given $k$--scale. Greater $k$'s correspond to
scales entering a non--linear regime earlier; therefore, we expect
greater discrepancies at greater $k$'s.

A test of these expectations can only be performed in a numerical way,
by running model simulations. Let us then outline that forthcoming
tomographic lensing experiments (see, e.g., \cite{RE})
will be able to measure spectra with a precision $\cal O$$(1\, \%)$
\cite{HT}, so that this is the limit above which spectral
discrepancies can no longer be disregarded. In turn, such experiments
are expected to test spectra up to $k \simeq 10\, h $Mpc$^{-1}$,
corresponding to length scales $\cal O$$(0.5\, h^{-1}$Mpc$)$, well
inside galaxy clusters. To test spectral regularities, then, hydro
effects cannot be disregarded.

A number of authors have considered these problems in some detail.
The $k$ value above which hydrodynamics causes spectral distorsions
$>1\, \%$ has been found to be $\sim 2$--$3\, h\, $Mpc$^{-1}$
\cite{JZ}. Models with an assigned polynomial DE state equation $w(z)$
(models $\cal A$) were then compared with models with constant $w$
(models $\cal W$, or {\it auxiliary} models), selected so to have the
same $L$ and the same $\sigma_8$ (mean square density fluctuation at
8$\, h^{-1}$Mpc), besides of $\Omega_{b,c}$ and $h$ \cite{FL}.
Spectral discrepancies, within $k \sim 2$--$3\, h\, $Mpc$^{-1}$ were
found to keep within $1\, \%$.

A critical feature of forthcoming mass probes is their capacity to
explore the Universe at $z > 0$. Models with the same conformal age at
$z=0$, and different DE state equation, have different ages at $z > 0$
and spectral discrepancies soon exceed $1\, \%$. In principle, then,
one should find an auxiliary model $\cal W$$(z)$ of $\cal A$, at any
$z$. But equal $H$ at $z \neq 0$ means different $H$ at $z=0$.
Accordingly, simulations performed in boxes of identical size in units
of $h_o^{-1}$Mpc, can no longer be compared (the $_o$ suffix is used
just here, to outline that $h$ is taken at $z=0$). If boxes of
suitable different sizes are then used, sample variance is the
dominant effect.

This difficulty was overcome by abandoning the request that $\cal A$
and $W(z)$ have the same $H(z)$, and requiring them to share just the
reduced density parameters $\omega_{c,b} =\Omega_{c,b} h^2$ {\it (weak
requirement)} \cite{CM}.
The definition of $\rho_{cr}$ (critical density) however implies that,
at any $z$,

\vskip .1truecm
$~~~~~~~~~~~~~~~~~~~~~~~~~~~~~~~~~~~
\Omega_{c,b} \times H^2 = \Omega_{c,b} \times (8\pi G/3) \rho_{cr}~,
$

\vskip .1truecm
\noindent
so that the $\omega_{c,b} \propto \Omega_{c,b} H^2 $ scales as $
\Omega_{c,b} \rho_{cr} = \rho_{c,b} \propto a^{-3}$, indipendently of
DE nature. Models with equal $\omega_{c,b}$ at $z=0$ will then share
them at any $z$. The $\cal W$$(z)$ model expected to approach $\cal A$
at $z$, then depends on $z$ just because of its different {\it
constant} DE state parameter $\tilde w(z)$. No constraint holds on
$H(z)$, any more; thus, we can take the same $H$ value in all models
at $z=0$.

Notice that the $\tilde w(z)$ dependence so defined is quite different
from $w(z)$, the DE state equation in $\cal A$. This outlines a
significant experimental risk, as observers fitting data against
constant $w$ models, would find $\tilde w(z)$, instead of the actual
DE state equation $w(z)$.

Here we extend the test in \cite{CM} to a set of models, sampling
cosmologies consistent with data \cite{K1,K2}, and confirm that the
weak requirement works up to the required $k$ (see also \cite{CM}). We
also test the requirement including hydrodynamics, confirming spectral
regularities up to $k \sim 10\, h$Mpc$^{-1}$ (see also \cite{CM1}).
Further hydrodynamical tests are in progress.

A side result we shall outline concerns the rate of star formation.
The size of the box used allows to follow such process in a rather
approximate way. We however tested that our simulations produce an
amount of star which is not unreasonable. Previous tests in boxes of
comparable side admittedly found more substantial difficulties
\cite{RZ}. The point is then that, in spite of the proximity of the
cosmologies treated, we find a significantly different star
production. If this property is confirmed by simulations in boxes
apt to follow in detail galaxy formation, it could envisage a new
important feature to trace DE nature observationally.

\section{N--body tests for cosmologies consistent with current data}
\begin{figure}[htbp]
\begin{center}
\vskip -.3truecm
\includegraphics[width=5.3cm]{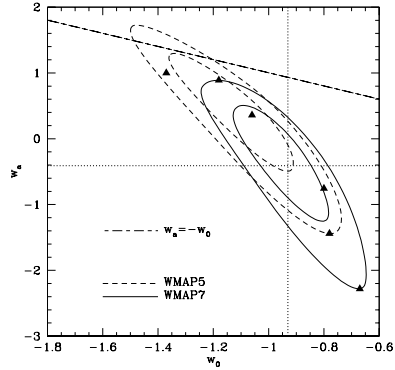}
\end{center}
\vskip -.3truecm
\caption{Likelihood ellipses on the $w_o$--$w_a$ plane from WMAP5 and
WMAP7 data releases. Black triangles indicate the $\cal A$ models
considered. State equations beyond the $w_o = -w_a$ line should be
modified at high--$z$. The dotted lines cross on the WMAP7 best fit.}
\label{ellipse2}
\end{figure}

Using available data, the WMAP team \cite{K1,K2} tried to constrain
the coefficients $w_0$ and $w_a$ in a polynomial expression $ w(a) =
w_0 + (1-a)\, w_a $ for the DE state parameter. However, the
likelihood ellipses (Figure \ref{ellipse2}) have significantly shifted
from WMAP5 to WMAP7.
%%%%%%%%%%%%%%%%%%%%%%%%%%%%%%%%%%%%%%%%%%%%%%%%%%%%%%%%%%%%%%%%%%%%%%
%%%%%%%%%%%%%%%%%%%%%%%%%%%%%%%%%%%%%%%%%%%%%%%%%%%%%%%%%%%%%%%%%%%%%%
%%%%%%%%%%%%%%%%%%%%%%%%%%%%%%%%%%%%%%%%%%%%%%%%%%%%%%%%%%%%%%%%%%%%%%
\begin{figure}[htbp]
\vskip -.5truecm
\begin{center}
\vskip -.3truecm
\includegraphics[width=10cm]{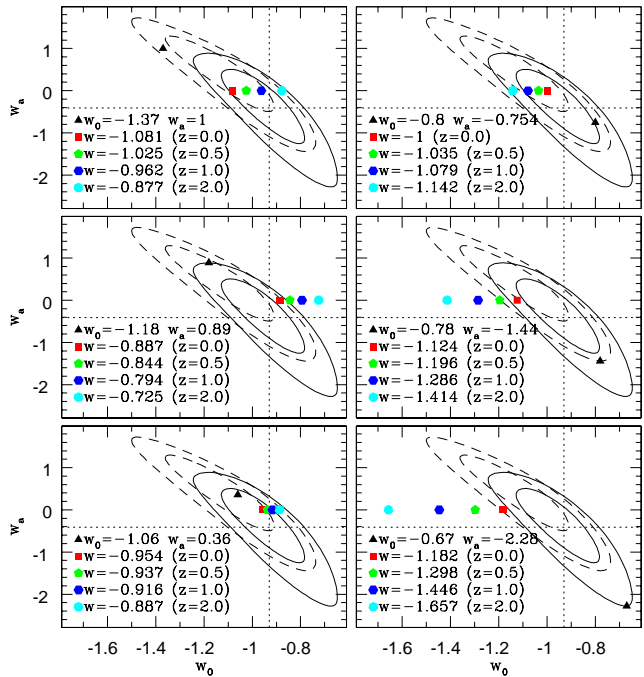}
\end{center}
\vskip -.8truecm
\caption{Each box refers here to a single $\cal A$ model in Figure
\ref{ellipse2} (black triangle) and related $\cal W$$(z)$ models
(color polygones). Notice that, in most cases, distances between
colored polygones are smaller than their distance from the black
triangle. This outlines the possibility of a serious bias in data
analysis, if data are tested by assuming constant--$w$ cosmologies.}
\label{Fe}
\end{figure}
%%%%%%%%%%%%%%%%%%%%%%%%%%%%%%%%%%%%%%%%%%%%%%%%%%%%%%%%%%%%%%%%%%%%%%
The setting of the six $\cal A$ models considered is shown in the
Figure. The other parameters are consistent with both WMAP5 and WMAP7
data: $\Omega_m=0.274$, $h=0.7$, $\sigma_8=0.81$ and $n_s=0.96$.

We test the spectra of $\cal A$ models {\it vs.} the corresponding
$\cal W$$(z)$. All simulations start at $z=24$, from realizations
fixed by an identical random seed. We use the {\sc pkdgrav}
code~\cite{stadel}, modified to deal with any variable $w(a)$, as in
\cite{CM}. In a box of side $L_{box}=256 h^{-1}$Mpc we set $N=256^3$
particles and use a gravitational softening $\epsilon = 25 h^{-1}
$kpc. We have 4 auxiliary models $\cal W$$(z)$ for $z = 0,~0.5,~1,~2~$
(see Fig.~\ref{Fe}), for each $\cal A$ model, run just down to the
redshift where they are tested. For all models, mass functions agree
with Sheth \& Tormen \cite{ST} predictions.

The first $\cal A$ cosmology considered, consistent with WMAP5, is
outside the 2--$\sigma$ curve for WMAP7. The significant shift of the
ellipses is not due to the fresh CMB inputs. Rather, omitting the
distance prior, as well as the whole SDSS data \cite{K1,K2}, surely
had an impact on it.

Figure \ref{Fe} shows that the distance between $\cal W$ models, on
the $w_a=0$ line by definition, is mostly smaller than their distance
from $\cal A$. Distances depend on $w_o$ and are smaller for the
central $w_o$ values. There is therefore a possible observational
problem, as outlined in the Introduction.

Model spectra are worked out by FFT--ing the matter density field,
computed on a regular grid $N_G\times N_G\times N_G$ (with $N_G =
2048$) via a Cloud in Cell algorithm.
%%%%%%%%%%%%%%%%%%%%%%%%%%%%%%%%%%%%%%%%%%%%%%%%%%%%%%%%%%%%%%%%%%%%%%
\begin{figure}[htbp]
\begin{center}
\vskip -.8truecm
\includegraphics[width=10.cm]{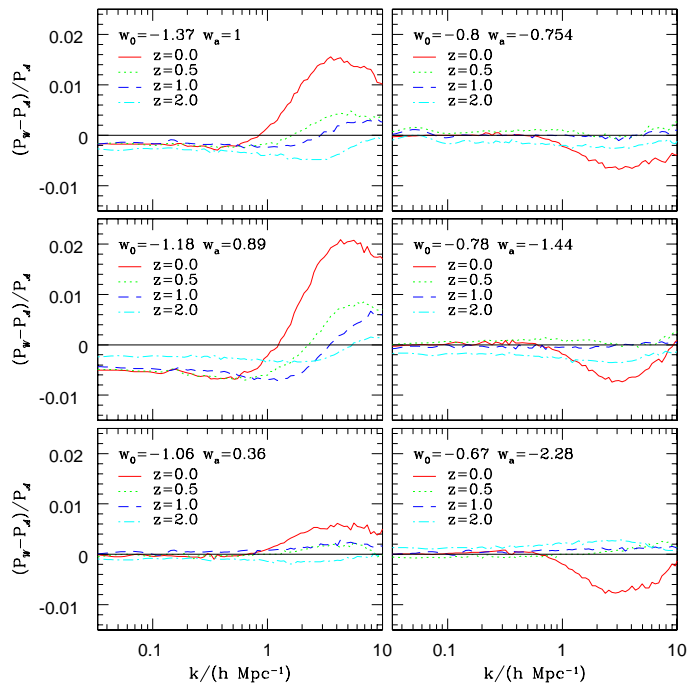}
\end{center}
\vskip -.4truecm
\caption{Spectral discrepancies at the redshift indicated by line
colors. Each box refers to an $\cal A$ model with the same ordering as
in Fig.~\ref{Fe}. Top discrepancies at $z=0$; only there they approach
$2 \, \%$ for some model with $w_o \ll -1$ and for $k \sim 4$--$5\,
h\, $Mpc$^{-1}$ (here however gas dynamics should not be ignored).
Within $\sim 2$--$3\, h\, $Mpc$^{-1}$, discrepancies keep $< 1\, \%~.$
}
\label{Fa}
\end{figure}
%%%%%%%%%%%%%%%%%%%%%%%%%%%%%%%%%%%%%%%%%%%%%%%%%%%%%%%%%%%%%%%%%%%%%%

Figure \ref{Fa} then shows spectral discrepancies. Within $k = 3\, h\,
$Mpc$^{-1}$ the top discrepancy occurs for the two models with lowest
$w_o$ ($<-1$), at $z=0$. Other models exhibit a nicer behavior,
allowing to guess that the greatest discrepancies, between $\cal A$
and $\cal W$$(z=0)$ evolutionary histories, occur for models with $w_o
\ll -1~.$ As expected, discrepancies decrease at higher $z$, where
non--linearities had no time to affect low $k$'s. At $z = 0.5$ they
are in the permil range, for all models.

\section{Hydrodynamical tests}
We then compare $\cal A$ and $\cal W$$(z)$ models, also including
baryon physics. Hydro simulations are much more time demanding and
here we present results for one model: $ w_o = -0.8, ~w_a = -0.754
\,$; top right in Figures \ref{Fe} and \ref{Fa}. The assigned
$\Omega_m$ value comes from $\Omega_b = 0.046, ~\Omega_c= 0.228$.

The program used is {\sc gasoline}, a multi-stepping, parallel
Tree--smoothed particle hydro (SPH) n--body code \cite{WS}.  It
includes radiative and compton cooling. The star formation, based on
\cite{KA}, allows gas particles above constant density and temperature
thresholds in convergent flows to spawn star particles at a rate
proportional to the local dynamical time \cite{SS}. The program
includes SN feedback \cite{SS} and a UV background, following
\cite{HM}. We applied {\sc gasoline} the same changes made in {\sc
pkdgrav} to handle dDE. Further details can be found in \cite{CM,GW}.

All simulations use $(2 \times) 256^3$ particles and most are run in a
box of 256$\, h^{-1}$Mpc. CDM (gas) particles have mass $m_c h/M_\odot
= 6.33 \times 10^{10}$ ($m_b h/M_\odot = 1.28 \times 10^{10}$). The
force resolution $\epsilon$ (softening) is 1/40 of the intra-particle
separation. For $256\, h^{-1}$Mpc this yields $\epsilon \simeq
25~h^{-1}$kpc (wavenumber $\kappa = 2\pi/\epsilon \simeq 150\, h\,
$Mpc$^{-1} $). All simulations start at $z = 24$. The $\cal A$ model
is selected so that $\cal W$$(z=0)$ is $\Lambda$CDM. This $\Lambda$CDM
was also run in a box of 64$\, h^{-1}$Mpc, still with $(2 \times)
256^3$ particles. All simulation parameters scale accordingly.

\begin{figure}
\vskip -.4truecm
\begin{center}
\includegraphics[width=5.5cm]{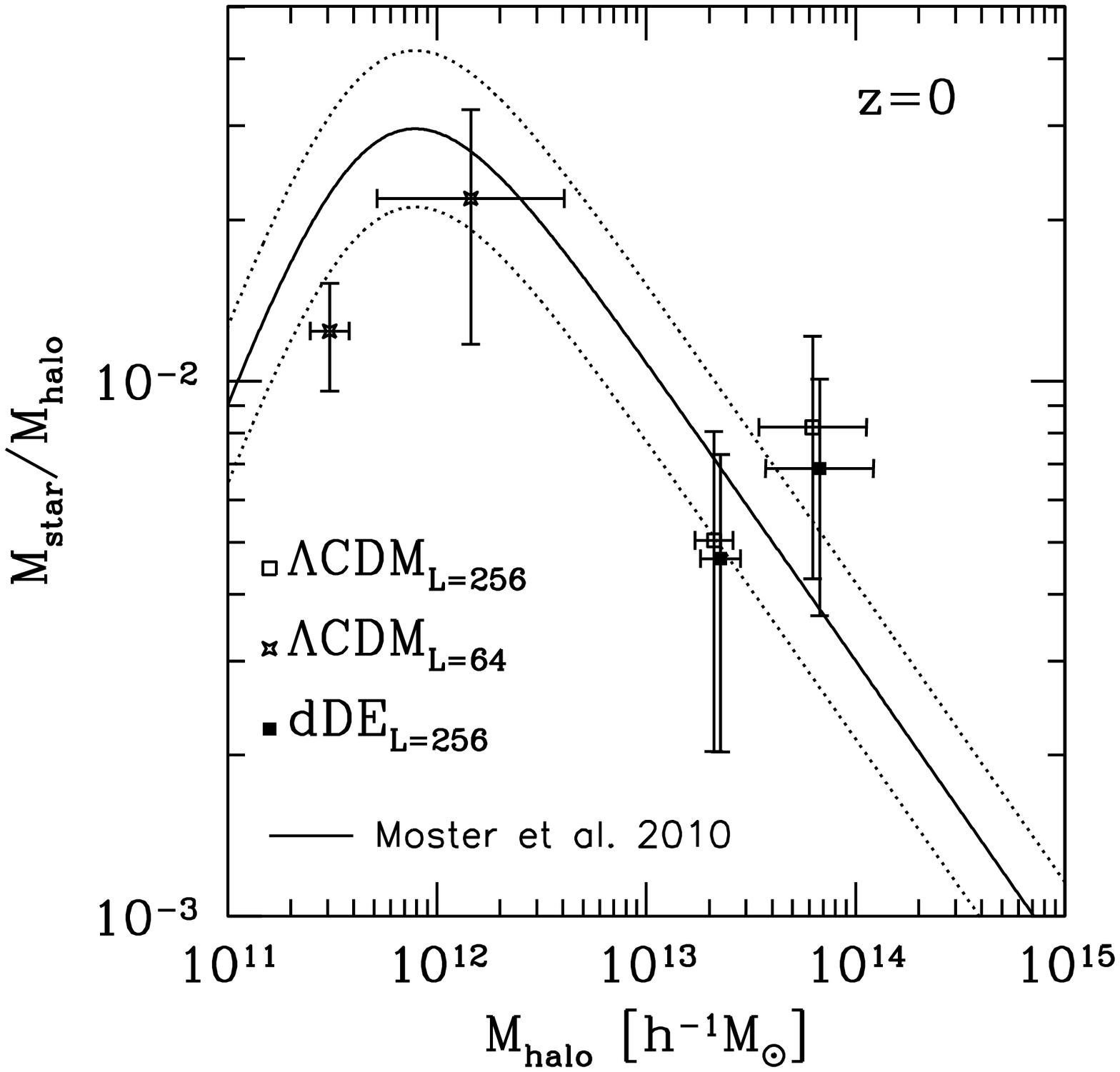}
\includegraphics[width=5.5cm]{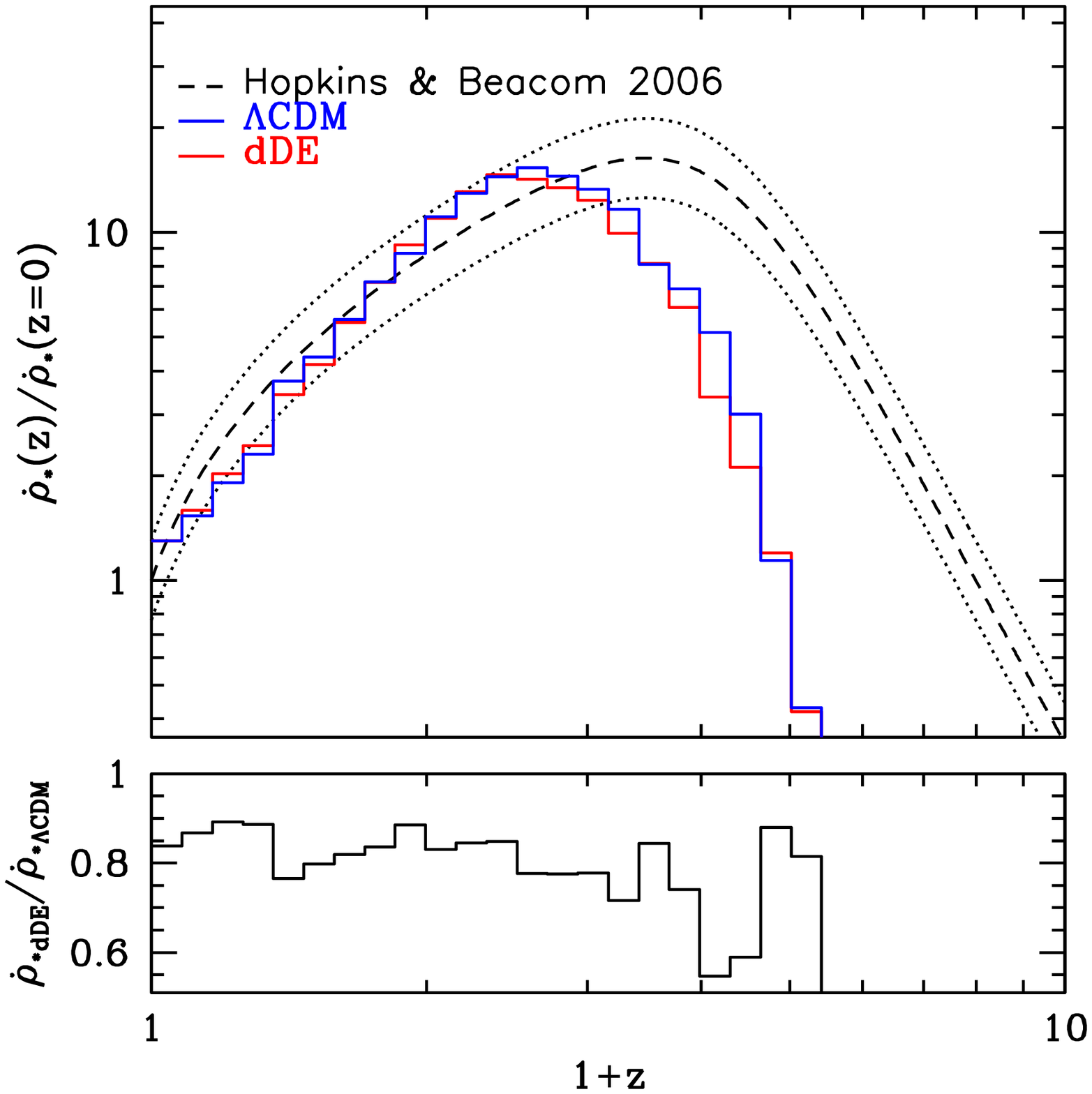}
\end{center}
\vskip -.4truecm
\caption{{\it L.h.s.}: Ratio between stellar and total mass in haloes.
Empty and solid squares show the results for the 256 box; crosses for
the 64 box. Results are compared with data in \cite{MS}; the 40\%
scatter yields $\sim \pm 2\sigma$. {\it R.h.s.}: Star formation
evolution in \LCDM and dDE in the 256 box; the dashed (black) line are
data fits \cite{HB}, assuming a Salpeter IMF; curves are normalized to
the $z=0$ value. The lower panel shows the ratio between star
productions in dDE and \LCDM.  }
\label{starhalo}
\end{figure}

Figure \ref{starhalo} (l.h.s.) shows the ratios $M_{star}/M_{halo}$
{\it vs.} $M_{halo}$ in both 64 and 256$\, h^{-1}$Mpc boxes, compared
with predictions in \cite{MS}.  The Figure shows that star formation
generally follows the halo occupation distribution trend. Slightly too
many stars form in haloes in the top $M_{halo}$ bin, due to
overcooling. This might be fixed by including AGN effects.
%%%%%%%%%%%%%%%%%%%%%%%%%%%%%%%%%      
%%%%%%%%%%%%%%%%%%%%%%%%%%%%%%%%%

The r.h.s. of Fig.~\ref{starhalo} shows star formation histories in
256$\, h^{-1}$Mpc boxes. Data \cite{HB} are also
shown, with $\pm 20\, \%$ error bars, approximating 2-$\sigma$'s. Our
outputs are shifted upwards by a factor 1.5$\, $, so to account for an
overall deficiency in star formation. Furthermore, observations and
simulations peak for $z \simeq 2.3\, $ and $\simeq 1.5:$ stars form
later in simulations. These discrepancies arise because stars do not
form in halos with $< 200$ gas particles \cite{CQ},
corresponding to a mass $\sim 1.3 \times 10^{12}\, h^{-1} M_\odot$ in
our simulations. In spite of that, at our resolution level star
formation seems described well enough to allow spectral comparison on
the scales relevant to us.

The bottom frame shows the ratio between star formation in $\cal A$
and $\Lambda$CDM. The decrease by $\sim 20 \%$ arises from a change of
$w(z),$ yelding a different evolutionary history, within an identical
conformal age. Future work will seek the cause of this unexpectedly
large shift.

\begin{figure}
\vskip -.5truecm
\begin{center}
\includegraphics[width=5.8cm]{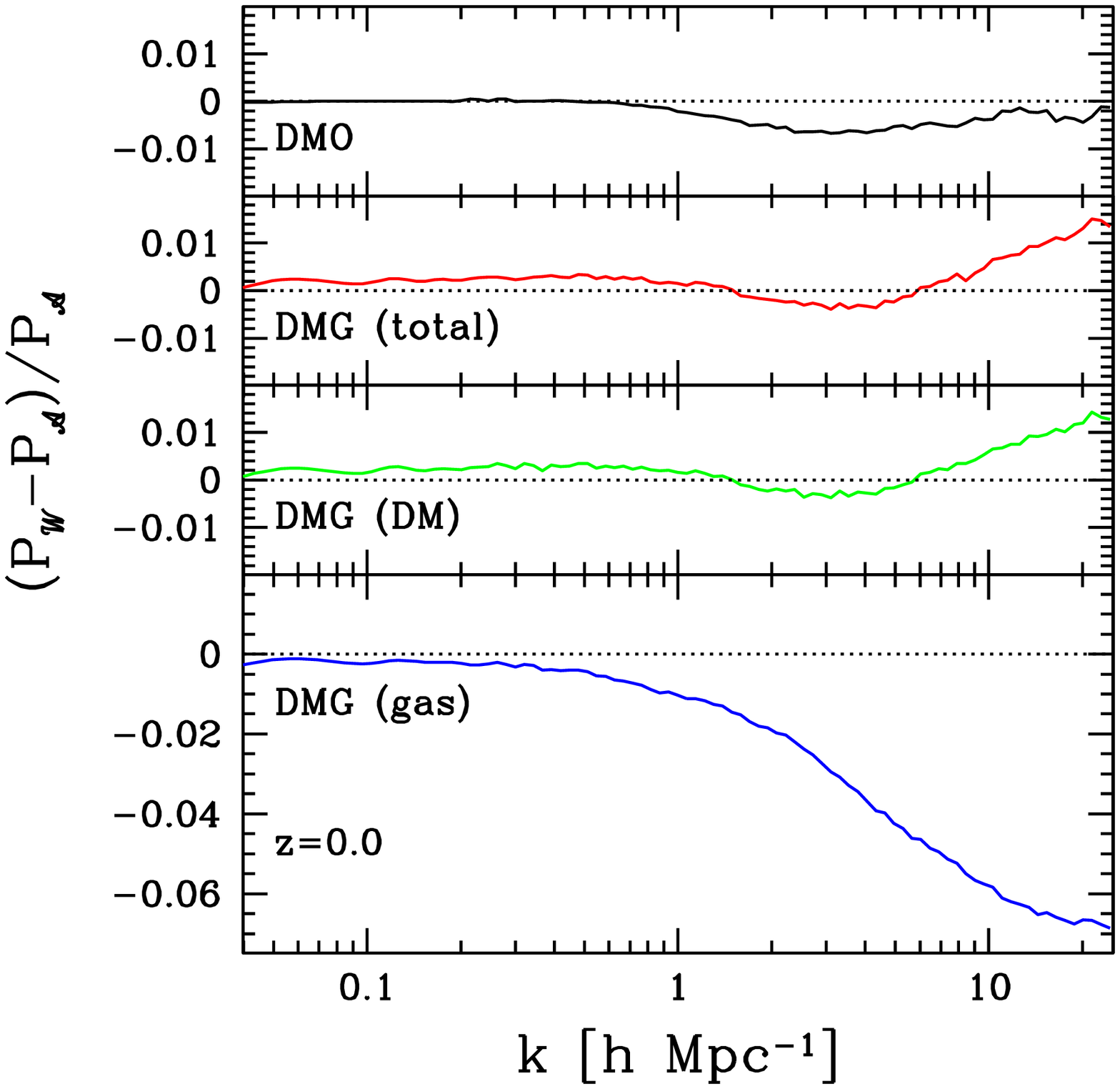}
\includegraphics[width=5.8cm]{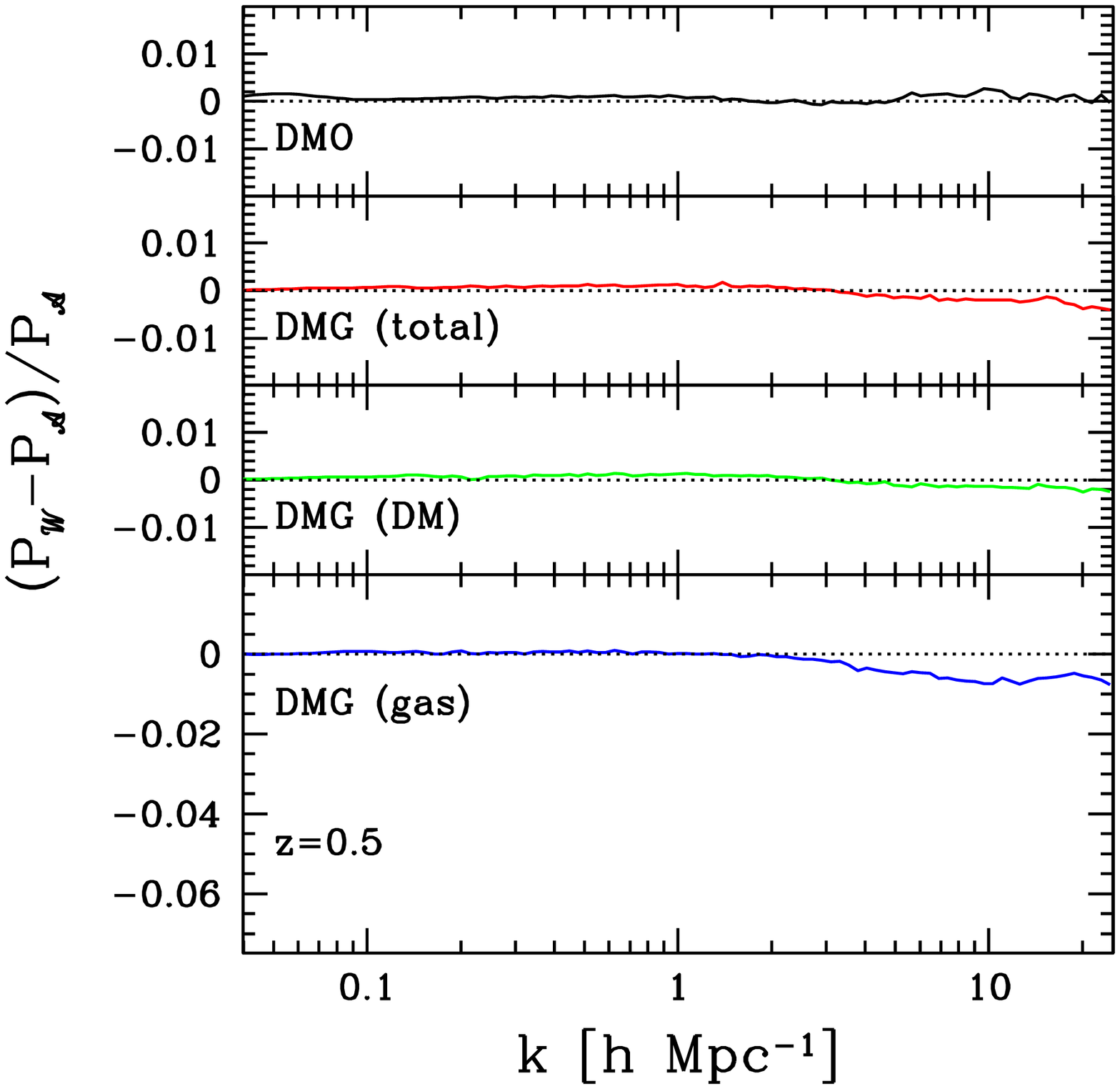}
\end{center}
\begin{center}
\includegraphics[width=5.8cm]{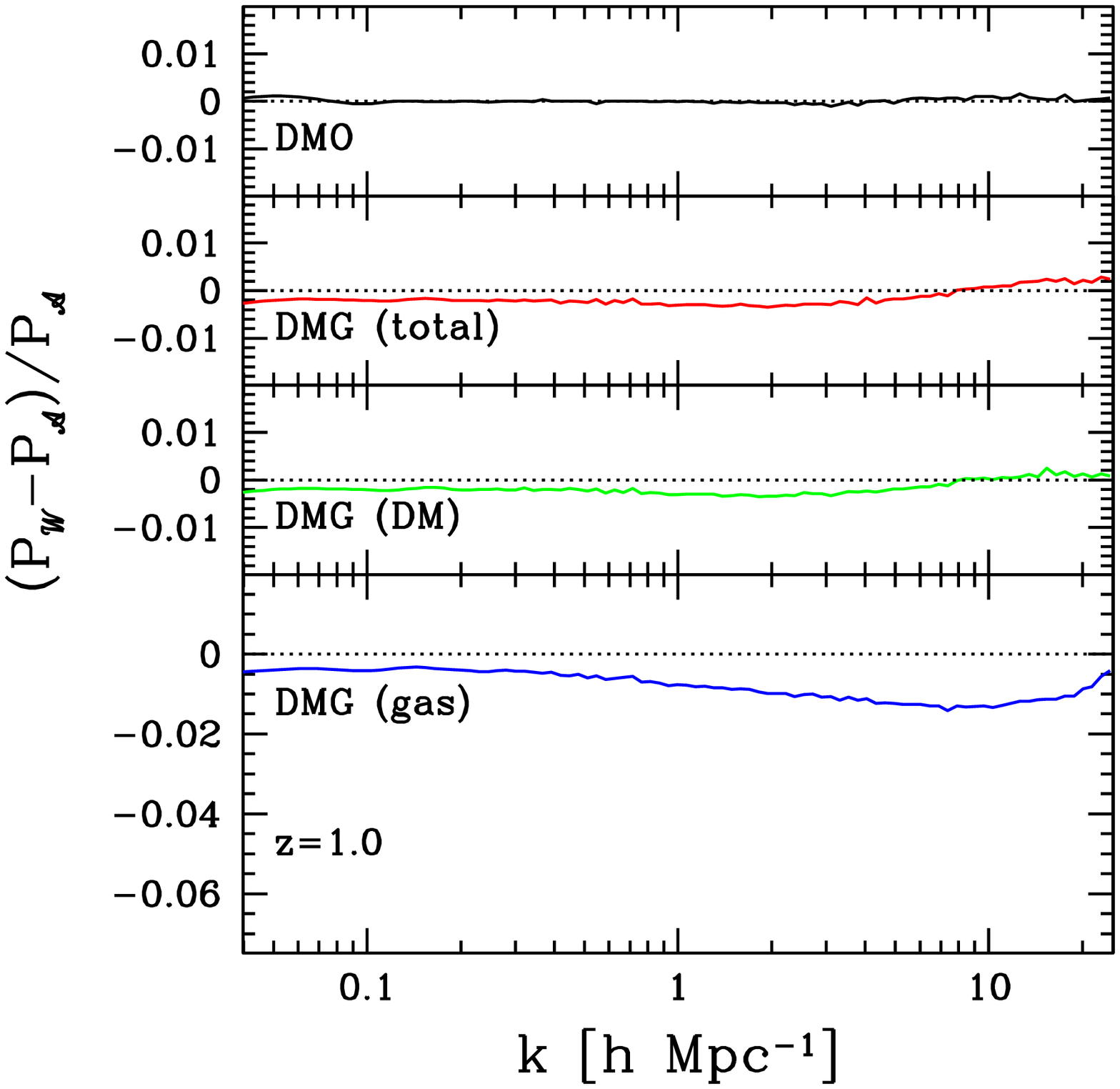}
\includegraphics[width=5.8cm]{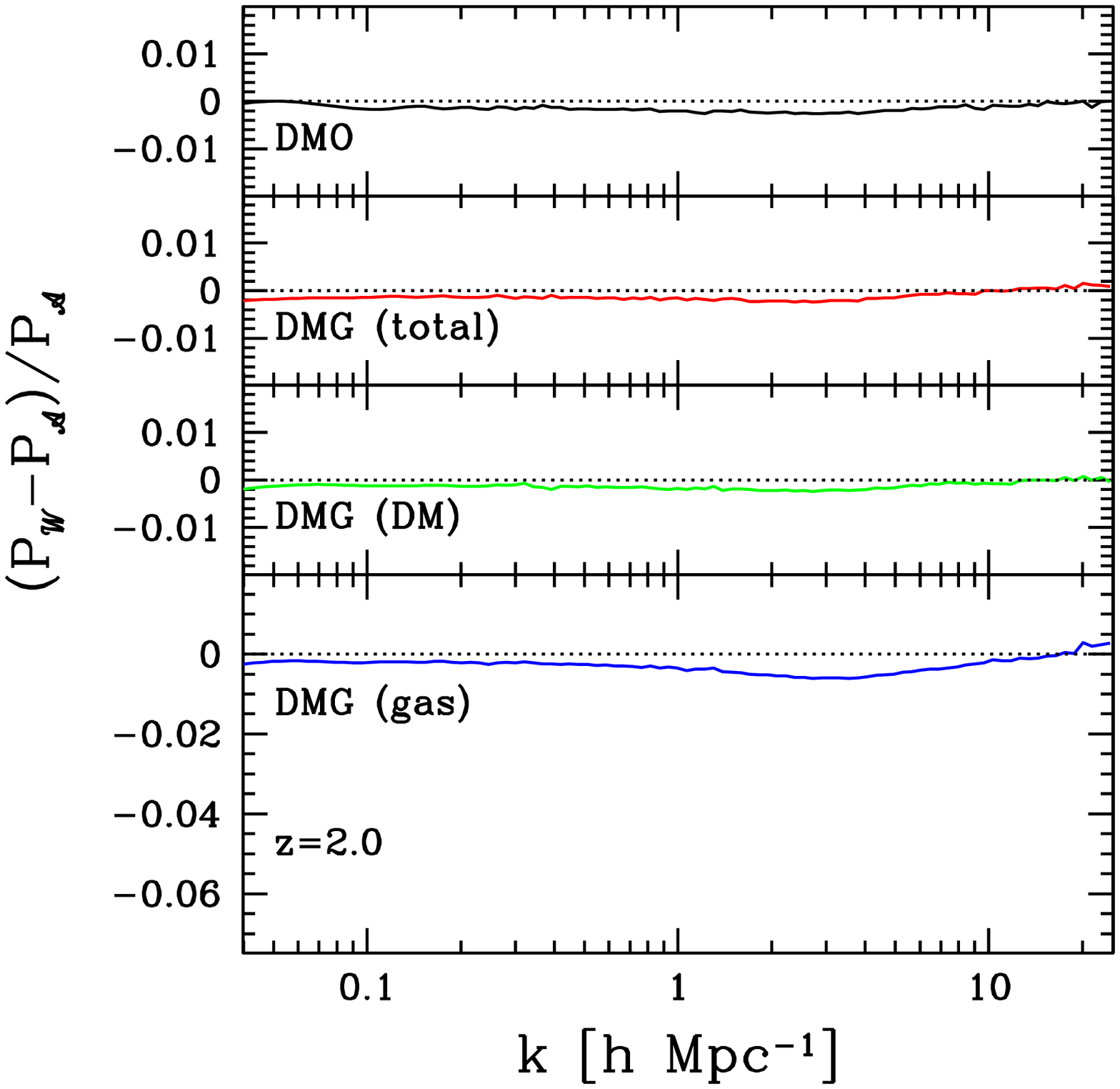}
\end{center}
\vskip -.5truecm
\caption{Comparison between $\cal A$ and $\cal W$$(z)$ spectra, for
n--body (DMO) and hydro (DMG) simulations (various components
shown). Top panels are for $z=0$ (l.h.s.) and $z=0.5$ (r.h.s.); bottom
panels for $z=1$ and $z=2$.}
\label{wr0a}
\end{figure}

Let us now come to spectral comparisons between $\cal A$ and $\cal
W$$(z)$ models at $z = 0$ and 0.5, 1, 2~. They are shown in Figures
\ref{wr0a}. 
%%%%%%%%%%%%%%%%%%%%%%%%%%%%%%%%%      
% WEAK REQUIREMENT % 
%-------------------------------
%%%%%%%%%%%%%%%%%%%%%%%%%%%%%%%%%      
For the model considered, regularities persist when gas dynamics is
important, although residual discrepancies (mostly of the order of a
few permils) are greater than in n--body simulations. The largest
discrepancy is found in the gas spectra, even if the gas spectrum
distortion at $z=0$, attaining $\sim 5\, \%$, apparently does not
imply a significant distortion in the global spectrum.

Altogether we conclude that even the inclusion of hydrodynamics keeps
the discrepancies between auxiliary and true model in the permil
range, for the model considered.

\end{document}